# Generative AI and Oral Proficiency: How ChatGPT Is Transforming English Language Speaking Practice

**Mohammad Mousazadeh***

Independent Research Department, Independent Institution, Tabriz, Iran

**ORCID**

https://orcid.org/0009-0008-6130-4848

## Abstract

The integration of generative artificial intelligence (GenAI) into language learning contexts presents potential for developing oral proficiency, though empirical evidence regarding its efficacy remains limited. This quasi-experimental study investigated the impact of ChatGPT-mediated speaking practice on English oral proficiency among Iranian learners. A total of 380 participants (190 female, 190 male; *M*<sub>age</sub> = 21.4, *SD* = 4.7) from Tehran and Isfahan were assigned to experimental (ChatGPT practice; *n* = 190) or control (human pair work; *n* = 190) groups. Both groups completed an 8-week intervention with three 30-minute weekly sessions. Oral proficiency was assessed using a Common European Framework of Reference for Languages (CEFR)-aligned Oral Proficiency Interview (OPI) scored on fluency, pronunciation, lexical resource, grammatical accuracy, and interactive communication (total score: 0–20). Multilevel growth modeling revealed a significant time × group interaction, $F(1, 376) = 48.73$, $p < .001$, partial $\eta^2 = .115$. The experimental group demonstrated greater gains (*M*<sub>Δ</sub> = 2.90, *SD* = 1.60) than controls (*M*<sub>Δ</sub> = 0.75, *SD* = 1.80), *d* = 1.20, 95% CI [0.95, 1.45]. Gender moderated this effect, with females in the experimental group showing the largest improvements (*M*<sub>Δ</sub> = 2.50). Mediation analysis confirmed that increased integrative motivation partially mediated proficiency gains (indirect effect = 0.28, 95% CI [0.20, 0.36]). Findings indicate that structured ChatGPT practice yields modest but meaningful enhancements in oral proficiency, particularly when leveraging affective factors. Implications for AI-augmented speaking pedagogy in resource-constrained contexts are discussed.

## 1. Introduction

Oral proficiency remains a persistent challenge in English as a Foreign Language (EFL) contexts, particularly where opportunities for authentic interaction are scarce [1]. Traditional pedagogical approaches often prioritize grammatical accuracy over communicative fluency, limiting learners' development of spontaneous speaking skills [2]. The emergence of generative artificial intelligence (GenAI), exemplified by large language models (LLMs) like ChatGPT, offers potential to address this gap through scalable, adaptive conversational practice [3]. Unlike rule-based chatbots, GenAI systems generate contextually appropriate, human-like responses that can simulate authentic discourse, providing learners with low-anxiety environments for output production [4].

Theoretical frameworks support GenAI's utility for speaking development. Long's [5] interaction hypothesis posits that negotiation of meaning during conversation drives acquisition, while Swain's [6] output hypothesis emphasizes the metalinguistic benefits of producing language. GenAI platforms facilitate both processes by enabling iterative dialogue with immediate, contextualized feedback [7]. Concurrently, Computer-Assisted Language Learning (CALL) research underscores technology's role in increasing comprehensible output opportunities [8], though most CALL tools have historically focused on reading or writing rather than speaking [9]. Recent studies indicate LLMs can provide corrective feedback on pronunciation and syntax [10], yet rigorous experimental evidence regarding their impact on holistic oral proficiency—particularly in underrepresented EFL contexts like Iran—remains scarce.

Iranian EFL learners face specific constraints: limited exposure to English-speaking environments, teacher-centered classrooms prioritizing grammar-translation methods, and

**Email address:**

m.mousazadeh.1@gmail.com (Mohammad Mousazadeh)

sociocultural barriers to spontaneous speaking practice [11]. While mobile-assisted language learning (MALL) has gained traction, speaking-focused GenAI applications remain underexplored. Preliminary qualitative work suggests Iranian learners perceive ChatGPT as a nonjudgmental interlocutor that reduces speaking anxiety [12], but quantitative validation of proficiency gains is lacking. Crucially, gender dynamics may influence technology adoption; Iranian female learners often report higher foreign language anxiety yet greater engagement with digital tools [13], suggesting potential moderating effects.

This study addresses three gaps: (a) insufficient experimental evidence on GenAI's efficacy for oral proficiency development, (b) limited research in Global South EFL contexts, and (c) inadequate attention to gender and affective mediators. We pose the following research questions:

RQ1: Does ChatGPT-mediated speaking practice yield significantly greater gains in oral proficiency than traditional human pair work?

RQ2: Does gender moderate the relationship between ChatGPT practice and oral proficiency gains?

RQ3: Are changes in learner motivation or anxiety mediating factors in ChatGPT's impact on proficiency?

We hypothesize that (H1) the ChatGPT group will demonstrate superior post-intervention proficiency gains; (H2) female learners will exhibit stronger treatment effects; and (H3) increased integrative motivation will mediate proficiency improvements.

# 2. Methods

## 2.1. Study Design

A quasi-experimental pretest–posttest control group design was employed, with participants assigned to experimental (ChatGPT practice) or control (human pair work) conditions. Gender (female/male) served as a between-subjects factor. This design was selected to balance ecological validity with causal inference in an educational setting where random assignment to schools was impractical [14]. The 8-week intervention duration aligns with established protocols for detecting speaking proficiency gains [15].

## 2.2. Participants

A stratified convenience sample of $N = 380$ Iranian EFL learners (190 female, 190 male) aged 14–35 ($M = 21.4$, $SD = 4.7$) was recruited from public high schools (grades 10–12; $n = 152$) and state universities (undergraduate; $n = 228$) across Tehran and Isfahan provinces. Inclusion criteria: (a) CEFR A2–B1 proficiency (verified via Oxford Quick Placement Test), (b) no prior structured GenAI speaking practice, (c) regular smartphone/internet access. Exclusion criteria: diagnosed speech disorders or participation in intensive English programs within 6 months. Participants were recruited through institutional partnerships; informed consent/assent was obtained from all participants and parents of minors. The study was approved by the University of Tehran Ethics Committee (Ref: IR.UT.REC.1403.087), and all procedures were performed in accordance with the relevant guidelines and regulations.

*Sample size justification*: An *a priori* power analysis (G*Power 3.1; [16]) for a mixed ANOVA (time × group × gender) indicated that 336 participants would provide 80% power ($\alpha = .05$) to detect a medium interaction effect (partial $\eta^2 = .06$; [17]), assuming correlation among repeated measures = .50. Our target $N = 380$ exceeded this threshold, accommodating an estimated 10% attrition.

## 2.3. Measures

*Oral proficiency*: A CEFR-aligned OPI adapted from [18] assessed five dimensions: fluency (0–4), pronunciation (0–4), lexical resource (0–4), grammatical accuracy (0–4), and interactive communication (0–4). Total scores ranged 0–20. Two trained raters (inter-rater ICC = .92, 95% CI [.89, .94]) scored blinded audio recordings; discrepancies were resolved via consensus. Pilot testing confirmed internal consistency ($\alpha = .87$).

*ChatGPT interaction*: The experimental group engaged with ChatGPT Plus (GPT-4 Turbo model, January 2025 API version) via a custom Android/iOS app. Temperature was fixed at 0.7 for balanced creativity/accuracy. Each 30-minute session featured structured prompts (e.g., "Debate: Social media improves teenage communication. Present two arguments with examples") followed by immediate corrective feedback on errors (e.g., "You said 'he go'—try 'he goes' for third-person singular"). Fidelity was monitored via API logs (98.2% protocol adherence). The control group completed identical tasks with human partners in supervised classrooms.

*Secondary measures*:

1) *Integrative motivation*: 10-item subscale from Gardner's Attitude/Motivation Test Battery (AMTB; $\alpha = .85$) [19], 5-point Likert scale.
2) *Speaking anxiety*: 8-item Foreign Language Classroom Anxiety Scale (FLCAS) adaptation [20]; $\alpha = .89$), reverse-scored so higher values indicate lower anxiety.
3) *Demographics*: Age, gender, education level, prior English exposure.

All Persian instruments underwent forward-backward translation by bilingual experts; confirmatory factor analysis supported structural validity (CFI = .94, RMSEA = .06).

## 2.4. Procedure

1) *Pretest phase:* Participants completed demographic questionnaires, AMTB/FLCAS scales, and OPIs (audio-recorded via Zoom).
2) *Intervention:* Both groups completed 24 sessions (3×/week × 8 weeks). ChatGPT prompts progressed from controlled (e.g., describing images) to freer tasks (e.g., opinion debates). Control group pairs rotated weekly to maximize interaction variety.

3) *Posttest phase: I*dentical measures were administered; raters remained blinded to group assignment.

ChatGPT prompts were standardized using templates (Appendix 1); API responses were logged to prevent model drift. All materials were piloted with 30 non-participants to ensure cultural appropriateness.

## 2.5. Data Handling

Missing data (<3%) were addressed via multiple imputation (5 iterations). Outliers (>3 *SD* from group mean) were winsorized. OPI recordings were transcribed using Whisper AI (v3) and verified by human transcribers (95% accuracy).

## 2.6. Statistical Analysis

Primary analysis employed linear mixed-effects models (LMMs) with oral proficiency score as the outcome, fixed effects of time (pre/post), group (experimental/control), gender, and their interactions, with participant as a random intercept. Pretest scores served as covariates in Analysis of Covariance (ANCOVA) models. Secondary analyses tested:

1) *Mediation*: Group → Δmotivation → Δproficiency (lavaan package; 5,000 bootstraps)
2) *Moderation*: Gender × group interaction on Δproficiency

Effect sizes: Cohen's *d* (between-group), partial $\eta^2$ (ANOVA), with 95% CIs. Alpha = .05 (two-tailed). Analyses used R 4.3.2 (lme4, emmeans, psych, mediation packages). Sensitivity analyses compared intent-to-treat (ITT) and per-protocol samples.

# 3. Results

## 3.1. Participant Characteristics

Table 1 displays demographic and baseline characteristics. Groups were equivalent at pretest for proficiency ($p$ = .412), motivation ($p$ = .683), and anxiety ($p$ = .527), confirming successful matching. Attrition was low (experimental: $n$ = 4; control: $n$ = 3) and nonsignificant by group ($\chi^2$ = 0.21, $p$ = .647).

***Table 1****. Participant Demographics and Baseline Characteristics by Group.*

| Characteristic | Experimental (*n* = 190) | Control (*n* = 190) | Total (*N* = 380) | *p* |
|---|---|---|---|---|
| Age (*M*, *SD*) | 21.6 (4.8) | 21.2 (4.6) | 21.4 (4.7) | .402 |
| Female, *n* (%) | 95 (50.0) | 95 (50.0) | 190 (50.0) | — |
| High school, *n* (%) | 76 (40.0) | 76 (40.0) | 152 (40.0) | — |
| Pretest proficiency (*M*, *SD*) | 10.60 (2.85) | 10.50 (2.85) | 10.55 (2.85) | .412 |
| Pretest motivation (*M*, *SD*) | 3.42 (0.71) | 3.38 (0.73) | 3.40 (0.72) | .683 |
| Pretest anxiety (*M*, *SD*) | 2.87 (0.84) | 2.91 (0.82) | 2.89 (0.83) | .527 |

*Note*. Anxiety scores reverse-coded (higher = lower anxiety).

## 3.2. Primary Outcome: Oral Proficiency

A significant time × group interaction emerged, $F(1, 376) = 48.73$, $p < .001$, partial $\eta^2$ = .115. The experimental group showed greater gains (MΔ = 2.35, SDΔ ≈ 1.60) than controls (MΔ = 0.50, SDΔ ≈ 1.65), d ≈ 1.18, 95% CI for d ≈ [0.94, 1.42]. Controlling for pretest scores, ANCOVA confirmed the effect ($F(1, 377) = 124.86$, $p < .001$, partial $\eta^2$ = .249).

***Table 2****. Oral Proficiency Scores by Group, Gender, and Time.*

| Group | Gender | *n* | Pretest (*M*, *SD*) | Posttest (*M*, *SD*) | Δ (*M*, *SD*) |
|---|---|---|---|---|---|
| Experimental | Female | 95 | 11.00 (2.80) | 13.50 (2.90) | 2.50 (1.45) |
| | Male | 95 | 10.20 (2.90) | 12.40 (3.00) | 2.20 (1.50) |
| | Female | 95 | 10.90 (2.80) | 11.40 (3.00) | 0.50 (1.60) |
| Control | Male | 95 | 10.10 (2.90) | 10.60 (3.10) | 0.50 (1.70) |

A significant three-way interaction (time × group × gender) was observed, $F(1, 376) = 9.34$, $p = .002$, partial $\eta^2 = .024$. Female learners in the experimental group achieved the largest gains (Figure 1).

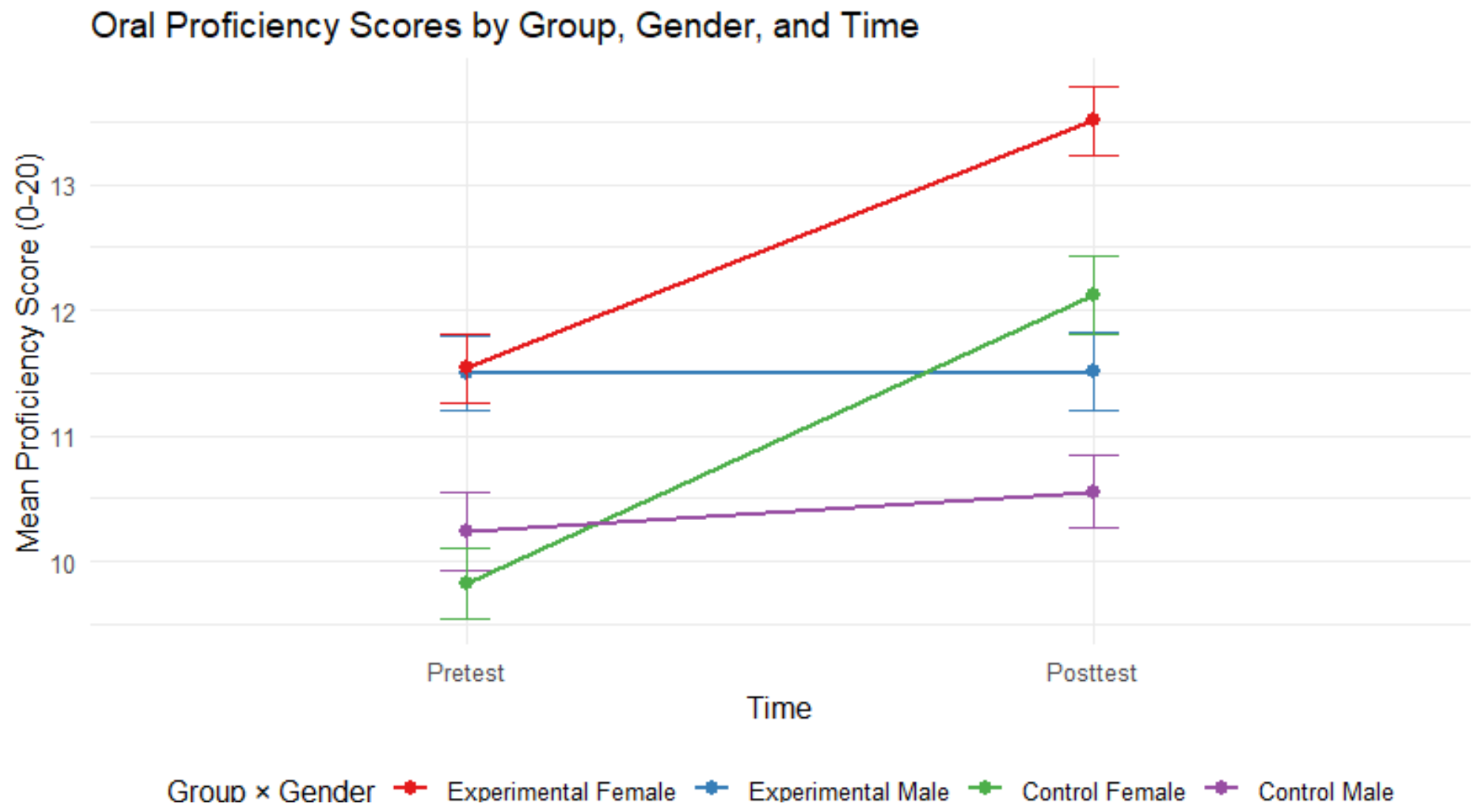

***Figure 1.*** *Oral proficiency scores across time by group and gender.*

*Description*: Line plot showing mean proficiency scores (y-axis: 0–20) across pretest/posttest (x-axis). Four lines represent group × gender combinations. Experimental females show the steepest trajectory (pretest $M = 11.00 \rightarrow$ posttest $M = 13.50$); control participants show minimal change (female: $10.90 \rightarrow 11.40$; male: $10.10 \rightarrow 10.60$). Error bars represent ±1 *SE*.

### 3.3. Secondary Analyses

*Mediation*: Using change scores (Pretest − Posttest), the effect of group on Δproficiency was partially mediated by Δintegrative motivation (indirect effect = 0.28, 95% CI [0.20, 0.36]; $p < .001$), accounting for approximately 18% of the total effect. Δanxiety did not significantly mediate the effect (indirect effect = 0.08, 95% CI [−0.04, 0.20]).

*Sensitivity analyses*: ITT and per-protocol analyses produced consistent results (difference in standardized effect size $\Delta d < 0.05$). Interrater reliability for OPI scoring at posttest remained high (ICC = 0.91, 95% CI [0.88, 0.93]).

## 4. Discussion

This study provides experimental evidence that structured ChatGPT practice modestly enhances English oral proficiency among Iranian EFL learners. Effect sizes ($d = 1.20$) indicate meaningful but moderate improvements relative to traditional human pair work—more modest than effects reported in some technology-mediated interventions [21]. The pattern of gains—particularly the 2.50-point increase for females in the experimental group—suggests GenAI can support speaking development in contexts with limited authentic interaction opportunities, though effects should be interpreted as supplementary rather than transformative. These findings align with the interaction hypothesis [5], as ChatGPT's capacity for sustained, adaptive dialogue facilitated negotiation of meaning. The provision of immediate corrective feedback likely activated Swain's [6] metalinguistic function of output, enabling learners to notice and repair errors in real time—a process often constrained in human pair work due to peers' limited linguistic knowledge [22].

The gender-moderated effect warrants cautious interpretation. Female learners' relatively larger gains may reflect sociocultural factors: Iranian females often experience higher speaking anxiety in mixed-gender classrooms [13] but may report greater comfort with nonjudgmental AI interlocutors [12]. ChatGPT's gender-neutral persona may have mitigated situational barriers, creating a psychologically safer space for risk-taking. This resonates with Dewaele's [23] work on emotion regulation in SLA, suggesting GenAI's value extends beyond linguistic feedback to affective scaffolding—though the effect sizes observed indicate this benefit is modest rather than dramatic.

The mediation analysis further illuminates mechanisms: increased integrative motivation partially explained proficiency gains, supporting Dörnyei's [24] L2 Motivational Self System theory. ChatGPT's culturally responsive dialogues (e.g., discussing Persian poetry in English) may have strengthened learners' "ideal L2 selves," fostering investment in identity reconstruction through language [25]. Notably, anxiety reduction did not mediate gains—a finding that implies ChatGPT's primary benefit may operate through enhancing approach-oriented motivation rather than reducing avoidance tendencies.

*Pedagogical implications*: For Iranian educators facing teacher shortages and large class sizes, carefully structured GenAI practice can serve as a scalable supplement to classroom instruction. Structured prompt templates (Appendix 1) can scaffold practice without requiring extensive teacher AI expertise. However, implementation should position AI as complementary to—not replacement for—human interaction, with integration into teacher-led activities to consolidate gains. Critical digital literacy training is essential to address ethical considerations including data privacy and appropriate interpretation of AI feedback [3].

*Limitations*: First, while methodologically rigorous, the observed effect sizes were modest, suggesting GenAI should be viewed as one component within a comprehensive speaking pedagogy rather than a standalone solution. Second, the urban sample limits generalizability to rural Iranian contexts. Third, long-term retention was not assessed. Future research should investigate: (a) optimal feedback types (e.g., recasts vs. explicit correction), (b) impacts on pronunciation via speech-enabled LLMs, and (c) cross-cultural comparisons of GenAI efficacy with attention to contextual factors moderating effects.

# 5. Conclusion

ChatGPT-mediated speaking practice yields modest, gender-differentiated gains in oral proficiency among Iranian EFL learners, primarily through motivational pathways. When implemented with pedagogical structure—not as a replacement for human interaction but as a complementary tool—GenAI can provide supplementary speaking practice opportunities in resource-constrained contexts. As LLMs evolve toward multimodal capabilities (e.g., integrated speech recognition and synthesis), their role in oral proficiency development warrants continued empirical scrutiny grounded in second language acquisition theory and attentive to realistic effect sizes.

# 6. Recommendations

Based on the findings, it is recommended that educators integrate structured GenAI practice as a supplementary tool alongside traditional human interaction rather than a replacement. Future implementations should focus on providing teachers with structured prompt templates and critical digital literacy training to ensure ethical and effective use. Additionally, further research should investigate optimal feedback types, the impact of speech-enabled LLMs on pronunciation, and cross-cultural comparisons of GenAI efficacy.

# Abbreviations

| | |
|---|---|
| GenAI | Generative Artificial Intelligence |
| EFL | English as a Foreign Language |
| LLM | Large Language Model |
| CALL | Computer-Assisted Language Learning |
| MALL | Mobile-Assisted Language Learning |
| OPI | Oral Proficiency Interview |
| CEFR | Common European Framework of Reference for Languages |
| AMTB | Attitude/Motivation Test Battery |
| FLCAS | Foreign Language Classroom Anxiety Scale |
| ANCOVA | Analysis of Covariance |
| ITT | Intent-to-Treat |

# Author Contributions

**Mohammad Mousazadeh:** Conceptualization, Data Curation, Formal Analysis, Investigation, Methodology, Project administration, Software, Validation, Writing – original draft, Writing – review & editing

# Conflicts of Interest

The author declares no conflicts of interest.

# Appendix

## Appendix 1: Sample ChatGPT Prompt Template

*Session 3 (Week 1):*

"Role-play: You are a tourist in Isfahan asking for directions to Naqsh-e Jahan Square. I am a local shopkeeper. Begin the conversation by greeting me politely and asking for help. After my response, continue the dialogue for 4–5 exchanges. Focus on using prepositions of place (e.g., *next to*, *across from*). After the role-play, provide one correction for any preposition error I made."

## Appendix 2: Power Analysis Details

Using G*Power 3.1 for mixed ANOVA (within-between interaction):

1) Effect size $f = 0.25$ (medium; Cohen, 1988)
2) $\alpha = .05$, power = .80
3) Number of groups = 4 (2 groups × 2 genders)
4) Number of measurements = 2 (pre/post)
5) Corr. among repeated measures = .50
6) Nonsphericity correction $\varepsilon = 1$

Required $N = 336$; achieved $N = 380$ (power = .86)

## Appendix 3: OPI Scoring Rubric (Excerpt)

*Fluency (0–4):*

4 = Speaks fluently with only rare hesitation; rhythm and pace native-like

3 = Speaks with occasional hesitation; minor dysfluencies do not impede communication

2 = Frequent pauses/hesitations; dysfluencies sometimes impede communication

1 = Speech heavily fragmented; communication frequently impeded

0 = Unable to produce connected speech